\documentstyle[12pt,twoside,fleqn,espcrc1]{article}

\begin{document}

% ---------- Gradient, etc.
\def\square{\hbox{{$\sqcup$}\llap{$\sqcap$}}}   % box
\def\grad{\nabla}                               % gradient
\def\del{\partial}                              % synonym for \partial

% ---------- Fractions.
\def\frac#1#2{{#1 \over #2}}
\def\smallfrac#1#2{{\scriptstyle {#1 \over #2}}}
\def\half{\ifinner {\scriptstyle {1 \over 2}}
   \else {1 \over 2} \fi}

% ---------- Bras and kets
\def\bra#1{\langle#1\vert}              % \bra{stuff} gives <stuff|
\def\ket#1{\vert#1\rangle}              % \ket{stuff} gives |stuff>

\def\bfgamma{\mbox{\boldmath$\gamma$}}

%       \simge and \simle make the "greater than about" and the "less
% than about" symbols with spacing as relations.
\def\simge{\mathrel{%
   \rlap{\raise 0.511ex \hbox{$>$}}{\lower 0.511ex \hbox{$\sim$}}}}
\def\simle{\mathrel{
   \rlap{\raise 0.511ex \hbox{$<$}}{\lower 0.511ex \hbox{$\sim$}}}}

%       \parenbar puts a bar in small parentheses over a character to
% indicate an optional antiparticle. \nunubar and \ppbar are special
% cases.

\def\parenbar#1{{\null\!                        % left-hand spacing
   \mathop#1\limits^{\hbox{\fiverm (--)}}       % (--) in 5pt type
   \!\null}}                                    % right-hand spacing
\def\nunubar{\parenbar{\nu}}
\def\ppbar{\parenbar{p}}

%       \buildchar makes a compound symbol, placing #2 above #1 and #3
% below it with \limits. \overcirc is a special case.

\def\buildchar#1#2#3{{\null\!                   % \null, cancel space
   \mathop#1\limits^{#2}_{#3}                   % #1, #2 above, #3 below
   \!\null}}                                    % cancel space, \null
\def\overcirc#1{\buildchar{#1}{\circ}{}}

%  \slashchar puts a slash through a character to represent contraction
%  with Dirac matrices. Use \not instead for negation of relations, and use
%  \hbar for hbar.

\def\slashchar#1{\setbox0=\hbox{$#1$}           % set a box for #1 
   \dimen0=\wd0                                 % and get its size
   \setbox1=\hbox{/} \dimen1=\wd1               % get size of /
   \ifdim\dimen0>\dimen1                        % #1 is bigger
      \rlap{\hbox to \dimen0{\hfil/\hfil}}      % so center / in box
      #1                                        % and print #1
   \else                                        % / is bigger
      \rlap{\hbox to \dimen1{\hfil$#1$\hfil}}   % so center #1
      /                                         % and print /
   \fi}                                         %

%       \subrightarrow#1 puts the text #1 under an arrow of the 
% appropriate length.

\def\subrightarrow#1{%                          % #1 under arrow
  \setbox0=\hbox{%                              % set a box
    $\displaystyle\mathop{}%                    % no mathop
    \limits_{#1}$}%                             % just limits
  \dimen0=\wd0%                                 % get width
  \advance \dimen0 by .5em%                     % add a bit
  \mathrel{%                                    % space like =
    \mathop{\hbox to \dimen0{\rightarrowfill}}% % arrow to width
       \limits_{#1}}}                           % text below

% ---------- Functions -- all defined like \sin, etc. in Plain TeX:
\def\real{\mathop{\rm Re}\nolimits}     % Re for real part
\def\imag{\mathop{\rm Im}\nolimits}     % Im for imaginary part

\def\tr{\mathop{\rm tr}\nolimits}       % tr for trace
\def\Tr{\mathop{\rm Tr}\nolimits}       % Tr for functional trace
\def\Det{\mathop{\rm Det}\nolimits}     % Det for functional determinant

\def\mod{\mathop{\rm mod}\nolimits}     % mod for modulo
\def\wrt{\mathop{\rm wrt}\nolimits}     % wrt for with respect to

% ---------- Abbreviations for units

\def\TeV{{\rm TeV}}                     % 10^12 electron volts
\def\GeV{{\rm GeV}}                     % 10^9  electron volts
\def\MeV{{\rm MeV}}                     % 10^6  electron volts
\def\KeV{{\rm KeV}}                     % 10^3  electron volts
\def\eV{{\rm eV}}                       % 1     electron volt

\def\mb{{\rm mb}}                       % 10^-27 cm^2
\def\mub{\hbox{$\mu$b}}                 % 10^-30 cm^2
\def\nb{{\rm nb}}                       % 10^-33 cm^2
\def\pb{{\rm pb}}                       % 10^-36 cm^2

%----------- References
%
%\def\journal#1#2#3#4{{\sl #1}\ $\underline{\hbox{#2}}$, {#3} ({#4})}
\def\journal#1#2#3#4{\ {#1}{\bf #2} ({#3})\  {#4}}
%     Specific journals:

\def\AdvPhys{\journal{Adv.\ Phys.}}
\def\AnnPhys{\journal{Ann.\ Phys.}}
\def\EurophysLett{\journal{Europhys.\ Lett.}}
\def\JApplPhys{\journal{J.\ Appl.\ Phys.}}
\def\JMathPhys{\journal{J.\ Math.\ Phys.}}
\def\LettNuovoCimento{\journal{Lett.\ Nuovo Cimento}}
\def\Nature{\journal{Nature}}
\def\NPA{\journal{Nucl.\ Phys.\ {\bf A}}}
\def\NPB{\journal{Nucl.\ Phys.\ {\bf B}}}
\def\NuovoCimento{\journal{Nuovo Cimento}}
\def\Physica{\journal{Physica}}
\def\PLA{\journal{Phys.\ Lett.\ {\bf A}}}
\def\PLB{\journal{Phys.\ Lett.\ {\bf B}}}
\def\PR{\journal{Phys.\ Rev.}}
\def\PRC{\journal{Phys.\ Rev.\ {\bf C}}}
\def\PRD{\journal{Phys.\ Rev.\ {\bf D}}}
\def\PRB{\journal{Phys.\ Rev.\ {\bf B}}}
\def\PRL{\journal{Phys.\ Rev.\ Lett.}}
\def\PhysRept{\journal{Phys.\ Repts.}}
\def\ProcNatlAcadSci{\journal{Proc.\ Natl.\ Acad.\ Sci.}}
\def\ProcRoySoc{\journal{Proc.\ Roy.\ Soc.\ London Ser.\ A}}
\def\RevModPhys{\journal{Rev.\ Mod.\ Phys. }}
\def\Science{\journal{Science}}
\def\SovPhysJETP{\journal{Sov.\ Phys.\ JETP }}
\def\SovPhysJETPLett{\journal{Sov.\ Phys.\ JETP Lett. }}
\def\SovJNuclPhys{\journal{Sov.\ J.\ Nucl.\ Phys. }}
\def\SovPhysDoklady{\journal{Sov.\ Phys.\ Doklady}}
\def\ZPhys{\journal{Z.\ Phys. }}
\def\ZPhysA{\journal{Z.\ Phys.\ A}}
\def\ZPhysB{\journal{Z.\ Phys.\ B}}
\def\ZPhysC{\journal{Z.\ Phys.\ C}}

\newcommand \beq{\begin{eqnarray}}
\newcommand \eeq{\end{eqnarray}}
\input epsf

%%%%%%%%%%%% titlepage %%%%%%%%%%%%%%%%%%%%%%%%%%%%%
\begin{titlepage}
\begin{flushright} {Saclay-T96/097}
\end{flushright}
\vspace*{1.5cm}
\begin{center}
\baselineskip=13pt
{\large {\bf  Lifetime of quasiparticles in hot gauge theories\\}}
\vskip0.5cm
Edmond Iancu\\
{\it Service de Physique Th\'eorique\footnote{Laboratoire de la Direction
des
Sciences de la Mati\`ere du Commissariat \`a l'Energie
Atomique}, CE-Saclay \\ 91191 Gif-sur-Yvette, France}\\

\vskip 2.cm
{\bf Abstract}
\end{center}

\begin{abstract}
The perturbative calculation of the lifetime of charged  excitations 
in ultrarelativistic plasmas is plagued with infrared
divergences which are not eliminated by the screening corrections.
  The physical processes responsible for these
divergences are the collisions involving the exchange of
longwavelength, quasistatic, magnetic gluons (or photons), which
are not screened by plasma effects.
In QED, the leading divergences can be resummed
 in a non-perturbative treatement based on
 a generalization of the Bloch-Nordsieck model at finite
temperature. The resulting expression of the fermion
 propagator is free of infrared problems, and exhibits
a {\it non-exponential} damping at large times:
$S_R(t)\sim
\exp\{-\alpha T \, t \ln\omega_pt\}$, where $\omega_p=eT/3$ is the plasma frequency
and $\alpha=e^2/4\pi$.
\end{abstract}

\vspace*{7.cm}

\begin{flushleft}

Contributed talk to ``Quark Matter'96'', the 12th International Conference
on Ultra-Relativistic Nucleus-Nucleus Collisions,
May 20-24, 1996, Heidelberg, Germany.\\
To appear in Nucl. Phys. {\bf A}
\end{flushleft}

\end{titlepage}

% put your own definitions here:
%   \newcommand{\cZ}{\cal{Z}}
%   \newtheorem{def}{Definition}[section]
%   ...
\newcommand{\ttbs}{\char'134}
\newcommand{\AmS}{{\protect\the\textfont2
  A\kern-.1667em\lower.5ex\hbox{M}\kern-.125emS}}

% add words to TeX's hyphenation exception list
\hyphenation{author another created financial paper re-commend-ed}

% declarations for front matter
\title{Lifetime of quasiparticles in hot gauge theories}

\author{Edmond Iancu}

\maketitle

\noindent{Service de Physique Th\'eorique
\footnote{Laboratoire de la Direction des
Sciences de la Mati\`ere du Commissariat \`a l'Energie Atomique}
, CE-Saclay, \\ 91191 Gif-sur-Yvette, France}

% typeset front matter

\vspace{.5cm}
\begin{abstract}
The perturbative calculation of the lifetime of charged  excitations 
in ultrarelativistic plasmas is plagued with infrared
divergences which are not eliminated by the screening corrections.
  The physical processes responsible for these
divergences are the collisions involving the exchange of
longwavelength, quasistatic, magnetic gluons (or photons), which
are not screened by plasma effects.
In QED, the leading divergences can be resummed
 in a non-perturbative treatement based on
 a generalization of the Bloch-Nordsieck model at finite
temperature. The resulting expression of the fermion
 propagator is free of infrared problems, and exhibits
a {\it non-exponential} damping at large times:
$S_R(t)\sim
\exp\{-\alpha T \, t \ln\omega_pt\}$, where $\omega_p=eT/3$ is the plasma frequency
and $\alpha=e^2/4\pi$.
\end{abstract}

\vspace{.5cm}
\section{INTRODUCTION}

The study of the elementary excitations of ultrarelativistic  plasmas,
such as  the quark-gluon plasma,  has received much
attention in the recent past.
(See \cite{BIO96,MLB96} for recent reviews and more references.)
The physical picture which emerges is that of a system with
 two types of degrees of freedom:
{\it i}) the plasma quasiparticles,
whose energy is of the order of the temperature $T$;
{\it ii}) the collective excitations, whose typical energy
is $gT$, where $g$ is the gauge coupling,
assumed to be small: $g\ll 1$ (in QED, $g=e$ is the electric charge).
For this picture to make sense, however, it is important that the
 lifetime of the excitations be large compared to the
typical period of the modes. 

Information about the lifetime is obtained from the
retarded propagator. A usual expectation is that
 $S_R(t,{\bf p})$ decays {\it exponentially} in time,
 $S_R(t,{\bf p})\,\sim\,{\rm e}^{-i E(p)t} {\rm e}^{ -\gamma({p}) t}$,
where $E(p) \sim T$ or $gT$ is the average energy of the excitation,
 and $\gamma(p)$ is the damping rate.
Therefore,  $|S_R(t,{\bf p})|^2\,\sim\,{\rm e}^{ -\Gamma({p}) t}$ 
with $\Gamma(p)=2\gamma(p)$, which identifies the lifetime
of the single particle excitation as $\tau(p) = 1/\Gamma(p)$.
The exponential decay may then be associated to a pole
of the Fourier transform  $S_R(\omega,{\bf p})$,
located at $\omega = E-i\gamma$.
The quasiparticles are well defined
if their lifetime  $\tau$ is much larger than the period $\sim 1/E$
of the field oscillations, that is, if the damping rate
$\gamma$  is small compared to the energy $E$. If this is the case,
 the respective damping rates
can be computed from the imaginary part of the on-shell
self-energy, $\Sigma(\omega=E(p), {\bf p})$. 

Previous calculations  \cite{Pisarski89} suggest that 
$\gamma\sim g^2T$
for both the single-particle and the collective excitations.
In the weak coupling regime $g\ll 1$,
 this is indeed small compared to the corresponding
energies (of order $T$ and $gT$, respectively),
suggesting that the  quasiparticles are well defined, and the
collective modes are weakly damped. However, the computation of 
$\gamma$ in perturbation theory 
is plagued with infrared (IR) divergences, which casts doubt on the
validity of these statements  [3---9].

The first attempts to calculate the damping rates
were made in the early 80's. It was then found that,
to one-loop order, the damping rate of the soft
collective excitations in the hot QCD plasma was gauge-dependent,
and could turn out negative in some gauges (see Ref. \cite{Pisarski91}
for a survey of this problem). Decisive progress on this 
problem was made by Braaten and Pisarski \cite{Pisarski89}
 who identified the resummation needed to obtain the
screening corrections in a gauge-invariant way 
(the resummation of the so called ``hard thermal loops'' (HTL)).
Such screening corrections are sufficient to
make IR-finite the transport cross-sections \cite{Baym90,BThoma91},
and also the damping rates of excitations 
with zero momentum \cite{Pisarski89,KKM}. 

At the same time, however,
it has been remarked \cite{Pisarski89} that the HTL resummation
 is not sufficient to render finite
 the damping rates of excitations with non vanishing momenta.
The remaining infrared divergences are due to collisions involving the
exchange of longwavelength, quasistatic, magnetic photons (or gluons),
which are not screened in the hard thermal loop approximation.
Such divergences affect the computation of the damping rates
of {\it charged} excitations (fermions and gluons),
 in both Abelian and non-Abelian gauge theories. 
Furthermore, the problem appears for both soft ($p \sim gT$) and hard 
 ($p \sim T$) quasiparticles. In QCD this problem is generally
 avoided by the {\it ad-hoc} introduction of an IR cut-off
(``magnetic screening mass'') $\sim g^2T$, which is
expected to appear dynamically from gluon
self-interactions \cite{MLB96}.
In QED, on the other hand, it is known that no magnetic
screening can occur \cite{Fradkin65},
so that the solution of the problem must lie somewhere else.

In order to make the damping rate $\gamma$ finite,
 Lebedev and Smilga proposed
a  self-consistent computation of  the damping rate,
by including $\gamma$  also in internal propagators \cite{Lebedev90}.
However, the resulting self-energy
is not analytic near the complex mass-shell, and the logarithmic
divergence actually reappears when the discontinuity of the self-energy is evaluated 
at $\omega= E - i\gamma$  \cite{Baier92,Pisarski93}. More
thorough investigations along the same lines led to the conclusion
that the full propagator has actually  no quasiparticle
pole in the complex energy plane \cite{Pilon93}. These analyses left
unanswered, however, the question of the large time
behavior of the retarded propagator.

As we have shown recently for the case of QED \cite{prl}, the answer
to this question requires a non perturbative treatment,
since infrared divergences occur in {\it all}
orders of  perturbation theory. We have identified the
 {\it leading} IR divergences in all orders, and
solved exactly an effective theory which reproduces all
these leading divergences. The resulting fermion
propagator $S_R(\omega)$ turns out to be {\it analytic}
in the vicinity of the mass-shell.
Moreover, for large times $t\gg 1/gT$,
the Fourier transform $S_R(t)$ does not show the usual exponential decay alluded
to before, but the more complicated behavior
$S_R(t)\,\sim\,{\rm e}^{-iE t} {\rm exp}\{-\alpha T \, t
\ln\omega_pt \}$, where  $\alpha=g^2/4\pi$ and $\omega_p\sim gT$ is the plasma frequency.
This corresponds to a typical lifetime $\tau^{-1}\sim g^2T\ln (1/g)$,
which is similar to the one provided by the perturbation
theory with an IR cut-off of the order $g^2T$.

\section{THE INFRARED PROBLEM}

Let me briefly recall how the infrared problem
occurs in the perturbative calculation of the damping rate $\gamma$.
For simplicity, I consider an Abelian plasma, as described by QED,
and compute the damping rate of a hard electron, with momentum $p\sim T$
and energy $E(p)=p$.

\begin{figure}
\protect \epsfxsize=8.cm{{\epsfbox{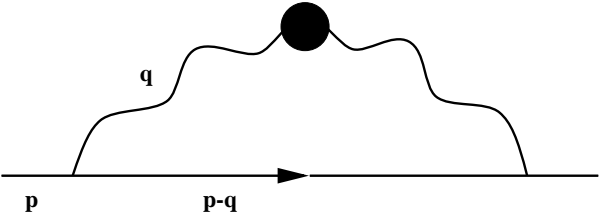}}}
	 \caption{The resummed one-loop self-energy}
\label{effS}
\end{figure}

\begin{figure}
\protect \epsfxsize=6.cm{\centerline{\epsfbox{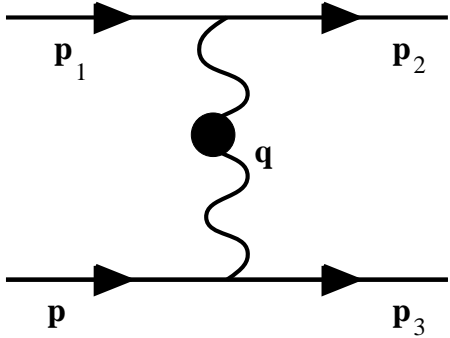}}}
	 \caption{Fermion-fermion elastic scattering in the Born approximation}
\label{Born}
\end{figure}

To leading order in $g$, and after the resummation of the screening 
corrections, $\gamma$ is obtained from the imaginary part of the 
effective one-loop self-energy in Fig.~\ref{effS}.
The blob on the photon line  in this figure
 denotes the  effective  photon propagator in the HTL approximation,
commonly  denoted as ${}^*D_{\mu\nu}(q)$. In the Coulomb gauge, the only non-trivial
components of ${}^*D_{\mu\nu}(q)$
 are the electric (or longitudinal)
one ${}^*D_{00}(q)\equiv {}^*\Delta_l(q)$, and the magnetic (or transverse) one
${}^*D_{ij}(q)=(\delta_{ij}-\hat q_i\hat q_j){}^*\Delta_t(q)$, with
 \beq\label{effd}
{}^*\Delta_l(q_0,q)\,=\,\frac{- 1}{q^2- \Pi_l(q_0,q)},\qquad
{}^*\Delta_t(q_0,q)\,=\,\frac{-1}{q_0^2-q^2 -\Pi_t(q_0,q)},\eeq
where $\Pi_l$ and $\Pi_t$ are the respective pieces
of the photon polarisation tensor \cite{BIO96,MLB96}.
Physically, the on-shell 
discontinuity of the diagram in Fig.~\ref{effS} accounts
for the scattering of the incoming electron (with four momentum
$p^\mu=(E(p), {\bf p})$) off a thermal fermion (electron or positron), 
as mediated by a soft, dressed, virtual photon. (See Fig.~\ref{Born}.)

The interaction rate corresponding to Figs.~\ref{effS} or \ref{Born}
 is dominated by soft momentum transfers $q\ll T$. It is
 easily computed as \cite{Pisarski93,BThoma91}
\beq\label{G2L}
\gamma \simeq\, \frac{g^4 T^3}{12}\,
\int_{0}^{q^*}{\rm d}q  \int_{-q}^q\frac{{\rm d}q_0}{2\pi}
\left\{|{}^*\Delta_l(q_0,q)|^2\,+\,\frac{1}{2}\left(1-\frac{q_0^2}{q^2}\right)^2
|{}^*\Delta_t(q_0,q)|^2\right\}\,,\eeq
where the upper cut-off $q^*$ distinguishes between
soft and hard momenta: $gT\ll q^* \ll T$.
Since the $q$-integral is dominated by IR momenta, its leading
order value is actually independent of $q^*$.

The two terms within the parentheses in eq.~(\ref{G2L})
  correspond to the exchange of an electric and of
a magnetic photon respectively.
For a bare (i.e., unscreened) photon, we have $|\Delta_l(q_0,q)|^2= 1/q^4$ and
$|\Delta_t(q_0,q)|^2= 1/(q_0^2-q^2)^2$, so that
 the $q$-integral in eq.~(\ref{G2L}) shows a quadratic IR divergence:
\beq\label{G2L0}
\gamma\simeq  \frac{g^4T^3}{8\pi} \,
\int_{0}^{q^*}\frac{{\rm d}q}{q^3}\,.\eeq
This divergence reflects the singular behaviour
of the Rutherford cross-section for forward scattering.
As well known, however, the quadratic divergence is removed by the
screening corrections contained in the photon polarization tensor.
We shall see below that the leading IR contribution
comes from the domain $q_0\ll q \ll T$, where we can
 use the approximate expressions \cite{BIO96,MLB96} (with $\omega_p=eT/3$)
\beq\label{pltstatic}
\Pi_l(q_0\ll q) \simeq  3{\omega_p^2}\,\equiv m_D^2,\qquad
\Pi_t(q_0\ll q) \simeq \,-i\,\frac{3\pi}{4}\,{\omega_p^2}\,\frac{q_0}{q}\,
.\eeq 
We see that screening occurs in different ways
in the electric and the magnetic sectors.
In the electric sector, the familiar static Debye screening provides
an IR cut-off $m_{D}\sim gT$. Accordingly,
 the electric contribution to $\gamma$ is finite,
and of the order  $\gamma_l \sim g^4 T^3/m_{D}^2
\sim g^2 T$. Its exact value can be computed by numerical
integration \cite{Pisarski93}. In the magnetic sector,
  screening occurs only for nonzero frequency $q_0$ \cite{Baym90}.
This comes from the imaginary part of the polarisation tensor,
and can be associated to the Landau damping \cite{PhysKin}
 of space-like photons ($q_0^2<q^2$).
This  ``dynamical screening'' is not sufficient to completely
remove the IR divergence of $\gamma_t\,$, which is only reduced to a logarithmic one:
\beq\label{G2LR}
\gamma_t &\simeq& \frac{g^4 T^3}{24}\,
\int_{0}^{q^*}{\rm d}q  \int_{-q}^q\frac{{\rm d}q_0}{2\pi}
\,\frac{1}{q^4 + (3\pi \omega_p^2 q_0/4q)^2} \nonumber\\
&\simeq & \frac{g^2T}{4\pi}
\int_{\mu}^{\omega_p}\frac{{\rm d}q}{q}\,=\, \frac{g^2T}{4\pi}\,
\ln \frac{\omega_p}{\mu}.\eeq
The unphysical lower cut-off $\mu$ has been introduced by hand,
in order to regularize the IR divergence of the integral over $q$.
The upper cut-off $\omega_p\sim gT$ 
 accounts approximately for the terms which have been
neglected when going from the first to the second line
of eq.~(\ref{G2LR}). As long as we are interested only in
the coefficient of the logarithm,
 the precise value of this cut-off is unimportant. The scale $\omega_p$ however
is uniquely determined by the physical process responsible for the existence
of space like photons, i.e., the Landau damping. As we
shall see later, this is the scale which fixes
the long time behavior of the retarded propagator.

The remaining IR divergence in eq.~(\ref{G2LR}) is due to collisions involving the
exchange of very soft ($q\to 0$),
 {\it quasistatic} ($q_0\to 0$) magnetic photons,
which are not screened by plasma effects.
To see that, note that the IR contribution to
 $\gamma_t$ comes from momenta $q\ll gT$,
where $|{}^*\Delta_t(q_0,q)|^2$ is almost a delta function of $q_0$:
\beq \label{singDT}
|{}^*\Delta_t(q_0,q)|^2\,\simeq\,
\frac{1} {q^4 + (3\pi \omega_p^2 q_0/4q)^2}\,
\longrightarrow_{q\to 0}\,\frac{4}{3 q \omega_p^2}\,\delta(q_0)\,.\eeq
This is so because,
as $q_0\to 0$, the imaginary part of the polarisation
tensor vanishes {\it linearly}
(see the second equation (\ref{pltstatic})), 
a property which can be related to the behaviour of the
phase space for the Landau damping processes.
Since energy conservation requires $q_0=q\cos\theta$, 
where $\theta$ is the angle
between the momentum of the virtual photon (${\bf q}$) and that
of the incoming fermion (${\bf p}$), 
the magnetic photons which are responsible for the singularity
are emitted, or absorbed, at nearly 90 degrees.

\section{A NON PERTURBATIVE CALCULATION}

\begin{figure}
\protect \epsfxsize=14.cm{\centerline{\epsfbox{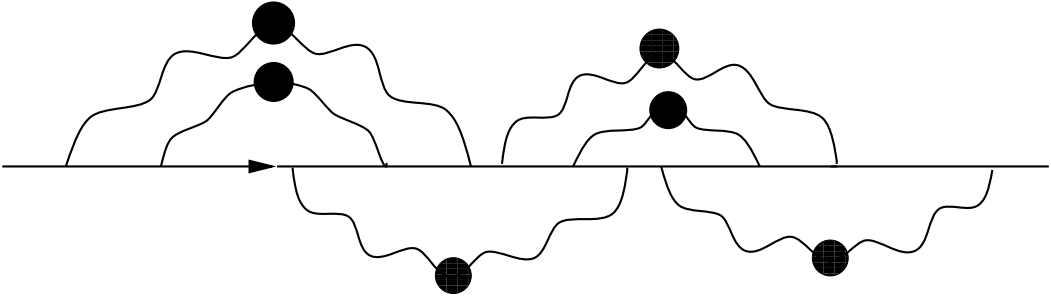}}}
	 \caption{A generic $n$-loop diagram (here, $n=6$)
for the self-energy in quenched QED.}
\label{effN}
\end{figure}

The IR divergence of the leading order calculation
invites to a more thorough investigation
of the higher orders contributions to $\gamma$. Such an
analysis \cite{prl} reveals strong, power-like, infrared
divergences, which signal the breakdown of the perturbation theory.
(A similar breakdown occurs in the computation of the 
corrections to the non-Abelian Debye mass \cite{debye}.)
To a given order in the loop expansion, the most
singular contributions to $\gamma$ arise from  self-energy diagrams
of the kind illustrated in Fig.~\ref{effN}. These diagrams have no
internal fermion loops (quenched QED), and all the internal  photons
are of the magnetic type (the electric photons, being screened,
give no IR divergences). Furthermore, the {\it leading} divergences arise,
in all orders, from the same kinematical regime as in the
one loop calculation,  namely from the regime where the
internal photons are soft ($q\to 0$) and quasistatic ($q_0\to 0$).
 This is so because of the specific IR behaviour
of the magnetic photon propagator, as illustrated in 
eq.~(\ref{singDT}). Physically, these divergences come
from multiple magnetic collisions.

This peculiar kinematical regime can be conveniently
exploited in the imaginary time formalism (see, e.g., \cite{MLB96}), 
where the internal photon lines carry only discrete
(and purely imaginary) energies, of the form $q_0=i \omega_n=
i 2\pi n T$, with integer $n$ (the so-called  Matsubara frequencies).
The non-static modes with $n\ne 0$ are well separated from
the static one $q_0=0$ by a gap of order $T$. I have argued before that
 the leading IR divergences come from the kinematical limit $q_0\to 0$.
Correspondingly, it can be verified \cite{prl} that, in the Matsubara formalism,
all these divergences are concentrated in diagrams in which
the photon lines are static, i.e., they 
 carry zero Matsubara frequency. (To one loop order,
this has been also notified in Refs. \cite{Marini92}.)
In what follows, we shall restrict
ourselves to these diagrams, and try to compute their contribution
to the fermion propagator near the mass-shell, in a non perturbative way.
Note that, for these diagrams, all the loop integrations 
are three-dimensional (they run over
the three-momenta of the internal photons), so that the associated IR divergences
are those of a three-dimensional gauge theory. This clearly
emphasizes the non perturbative character of the leading IR structure.

As we shall see now, this
``dimensional reduction'' brings in simplifications which allows one
 to arrive at an explicit solution of the problem \cite{prl}.
The point is that three-dimensional quenched QED can be
``exactly'' solved in the Bloch-Nordsieck approximation \cite{Bogoliubov},
which is the relevant approximation for the infrared structure
of interest. Namely, since the incoming fermion is interacting
only with very soft ($q\to 0$) static ($q_0=0$) magnetic photons,
its trajectory is not significantly deviated by the
successive collisions, and its spin state does not change.
This is to say, we can ignore the spin degrees of freedom,
which play no dynamical role, and we can assume the fermion to
move along a straightline trajectory with constant velocity
${\bf v}$ (for the ultrarelativistic hard  fermion,
$|{\bf v}|=1$; more generally, for the soft excitations,
${\bf v}(p) \equiv \del E(p)/\del {\bf p} = v(p) {\bf \hat p}$
 is the corresponding group velocity, with $|v(p)|< 1$).
Under these assumptions, the fermion propagator can be easily
computed as \cite{prl}
\beq\label{SRT}
S_R(t,{\bf p})&=&i\,\theta(t) {\rm e}^{-iE(p)t}\,\Delta(t),\eeq
where
\beq\label{SR0}
\Delta(t)=  {\rm exp}\left \{-g^2T
\int^{\omega_p} \frac{{\rm d}^3q}{(2\pi)^3} 
\,\frac{1}{q^2}\,\frac
{1-  {\rm cos}\,t ({\bf v}(p) \cdot {\bf q})}{
({\bf \hat p \cdot q})^2} \right\},\eeq
contains all the non-trivial time dependence. 
The integral in eq.~(\ref{SR0}) is formally identical to that one would 
get in the Bloch-Nordsieck model in 3 dimensions.
Note, however, the upper cut-off $\omega_p \sim gT$, which occurs for
the same reasons as in eq.~(\ref{G2LR}). Namely, it reflects the dynamical
cut-off at momenta $\sim gT$, as provided by the Landau damping.

The integral over $q$ has no infrared divergence, but one can verify 
that the expansion of $\Delta(t)$ in powers of $g^2$ generates 
the most singular pieces of the usual perturbative expansion
for the self-energy \cite{prl}.
Because our approximations preserve only
 the leading infrared behavior of the perturbation theory,
eq.~(\ref{SR0}) describes  only the leading {\it large-time} behavior
of $\Delta(t)$. Since the only energy scale in the momentum integral of
eq.~(\ref{SR0}) is the upper cut-off, of order $gT$,
 the large-time regime is achieved for $t\gg 1/gT$.
Note that, strictly speaking, eq.~(\ref{SR0}) holds only in the Feynman gauge.
However, its leading large time behaviour ---  
which is all what we can trust anyway ! ---
 is actually gauge independent  \cite{prl} and of the form (we set here $\alpha
= g^2/4\pi$ and  $v(p)=1$ to simplify writing)
\beq\label{DLT}
\Delta(\omega_pt\gg 1)\,\simeq \,{\rm exp}\Bigl( -\alpha Tt \ln \omega_p
t\Bigr).\eeq
 A measure of the decay time $\tau$ is given by 
\beq \frac{1}{\tau}=\alpha T\ln \omega_p \tau=
\alpha T\left(\ln \frac{\omega_p}{\alpha T} - \ln
\ln \frac{\omega_p}{\alpha T} + \,...\right).\eeq
Since $\alpha T \sim g\omega_p$,  $\tau \sim
 1/(g^2 T \ln (1/g))$. This corresponds to a damping rate 
$\gamma\sim1/\tau\sim g^2 T\ln (1/g)$, similar to that obtained in a
one loop calculation  with an IR cut-off
$\mu \sim g^2T$ (cf. eq.~(\ref{G2LR})).

However, contrary to what perturbation theory predicts, 
$\Delta(t)$ is decreasing faster than any exponential. It follows that
 the  Fourier transform 
\beq\label{SRE}
S_R(\omega, {\bf p})\,=\,
\int_{-\infty}^{\infty} {\rm d}t \,{\rm e}^{-i\omega t}
S_R(t,{\bf p})\,=\,
i\int_0^{\infty}{\rm d}t
\,{\rm e}^{it(\omega- E(p)+i\eta)}\,\Delta(t),\eeq
  exists 
for {\it any} complex (and finite) $\omega$. Thus,  the retarded propagator
 $S_R(\omega)$ is an entire function, with sole singularity at 
Im$\,\omega\to -\infty$.   The associated spectral density
$\rho(\omega, p)$ (proportional to the imaginary part of
$S_R(\omega, {\bf p})$) retains the shape
of a {\it resonance}  strongly peaked around the perturbative mass-shell 
$\omega = E(p)$, 
with a typical width of order $\sim g^2T \ln(1/g)$ \cite{prl}.

\section{CONCLUSIONS}

The previous analysis, and, in particular, the last conclusion
about the resonant shape of the spectral density,
confirm that the quasiparticles are well defined,
even if their mass-shell properties cannot be computed
in perturbation theory. The infrared divergences occur because
of the degeneracy between the mass shell of the charged
particle and the threshold for the emission or the absorbtion
of $n$ ($n \ge 1$) static transverse photons.
Note that the emitted photons are virtual, so,
strictly speaking, the physical
processes that we have in mind are the collisions between
the charged excitation and the thermal particles, with
the exchange of quasistatic magnetic photons.
The resummation of these multiple collisions to all orders
in $g$ modifies the analytic structure of the fermion
propagator and yields an unusual, non-exponential,
damping in time.

This result solves the IR problem of the damping rate 
in the case of QED. Since a similar problem occurs in QCD
as well, it is natural to ask what is the relevance
of the present solution for the non-Abelian plasma.
It is generally argued  --- and also supported by lattice
computations \cite{Tar} --- that the self-interactions
of the chromomagnetic gluons 
may generate  magnetic screening at the scale $g^2 T$
(see \cite{MLB96} and Refs. therein).
As a crude model, we may include a screening mass $\mu\sim g^2T$
in the magnetostatic propagator in the 
QED calculation. This amounts to replacing $1/q^2 \to
1/(q^2 + \mu^2)$ for the photon propagator in eq.~(\ref{SR0}).
After this replacement, the latter equation provides,
 at very large times $t\simge 1/g^2T$,
an exponential decay:  $\Delta(t)
\sim \exp(-\gamma t)$, with $\gamma = \alpha T\ln(\omega_p/\mu)
=  \alpha T\ln(1/g)$.
However, in the physically more interesting regime of intermediate
times $1/gT \ll t \ll 1/g^2 T$, the behavior is governed
uniquely by the plasma frequency, according to our result
(\ref{DLT}): $\Delta(t)\sim \exp ( -\alpha Tt \ln \omega_p
t)$.  Thus, at least within this limited model,
which is QED with a ``magnetic mass'', the time behavior
in the physical regime remains controlled by the
Bloch-Nordsieck mechanism. But, of course, this result gives no
serious indication about the real situation in QCD, since
it is unknown whether, in the present problem,
 the effects of the gluon self-interactions
can be simply summarized in terms of a magnetic mass.

To conclude, the results of Refs. \cite{prl,debye} suggest
that the infrared divergences of the ultrarelativistic
Abelian plasmas can be eliminated by soft photon resummations,
\`a la Bloch-Nordsieck. For non-Abelian plasmas, on the other hand,
much work remains to be done, and this requires, in particular,
the understanding of the non-perturbative sector of the
magnetostatic gluons.


\begin{thebibliography}{9}

\bibitem{BIO96}
 J.P. Blaizot, J.-Y. Ollitrault and E.~Iancu, Collective
Phenomena in the Quark-Gluon Plasma, in  Quark-Gluon Plasma 2,
Editor R.C. Hwa,  World Scientific, 1996.

\bibitem{MLB96}
M. Le Bellac, Recent Developments in Finite Temperature Quantum
 Field Theories,
Cambridge University Press, 1996.

\bibitem{Pisarski89}
R.D.~Pisarski, \PRL{63}{1989}{1129};
E.~Braaten and R.D.~Pisarski, \PRL{64}{1990}{1338}; 
 \PRD{42}{1990}{2156}; \NPB{337}{1990}{569}.
 
\bibitem{Lebedev90}
V.V Lebedev and A.V. Smilga, \PLB{253}{1991}{231}; \AnnPhys{202}{1990}{229};
Physica {\bf A181} (1992) 187.

\bibitem{Marini92}
C.P. Burgess and A.L. Marini,
 \PRD{45}{1992}{R17}; A.K. Rebhan, \PRD{46}{1992}{482};
F. Flechsig, H. Schulz and A.K. Rebhan,
 \PRD{52}{1995}{2994}.

\bibitem{Baier92}
R. Baier, H. Nakkagawa and A. Ni\'egawa, Can. J. Phys.
{\bf 71} (1993) 205.


\bibitem{Pisarski93}
R.D. Pisarski, \PRD{47}{1993}{5589}.

\bibitem{Altherr93}
%R.D. Pisarski, \PRD{47}{1993}{5589};
T. Altherr, E. Petitgirard and T. del Rio Gaztelurrutia, 
\PRD{47}{1993}{703}; H.~Heiselberg and C.J.~Pethick, \PRD{47}{1993}{R769};
%A.V. Smilga, Phys. Atom. Nuclei {\bf 57} (1994) 519;
A. Ni\'egawa, \PRL{73}{1994}{2023};
K. Takashiba, preprint hep-ph/9501223 (unpublished).


\bibitem{Pilon93}
S. Peign\'e, E. Pilon and D. Schiff, 
Z. Phys. {\bf C60} (1993) 455;
A.V. Smilga, Phys. Atom. Nuclei {\bf 57} (1994) 519;
R. Baier and R. Kobes, \PRD{50}{1994}{5944}.

\bibitem{Pisarski91}
 R.D.~Pisarski, \NPA{525}{1991}{175}.

\bibitem{Baym90}
G.~Baym, H.~Monien, C.J.~Pethick, and D.G.~Ravenhall,\PRL{64}{1990}{1867}.

\bibitem{BThoma91}
E.~Braaten and M.H.~Thoma, \PRD{44}{1991}{1298}.

\bibitem{KKM}
R.~Kobes, G.~Kunstatter and K.~Mak, \PRD{45}{1992}{4632}; E.~Braaten and
R.D.~Pisarski, \PRD{46}{1992}{1829}.

\bibitem{Fradkin65}
E. Fradkin, Proc. Lebedev Phys. Inst. {\bf 29} (1965) 7;
J.P. Blaizot, E. Iancu and R. Parwani, \PRD{52}{1995}{2543}.


\bibitem{prl}
J.P. Blaizot and E. Iancu, \PRL{76}{1996}{3080}
and hep-ph/9607303, to appear in Phys. Rev. {\bf D}.

\bibitem{PhysKin}
E.M.~Lifshitz and L.P.~Pitaevskii,  Physical Kinetics, Pergamon Press,
Oxford, 1981.

\bibitem{debye}
J.P. Blaizot and E. Iancu, \NPB{459}{1996}{559}.


\bibitem{Bogoliubov}
N.N.~Bogoliubov and D.V.~Shirkov,  Introduction to the Theory of
Quantized Fields, Interscience Publishers,  New-York, 1959.

\bibitem{Tar}
C. DeTar, Quark-Gluon Plasma in Numerical Simulations of Lattice QCD,
in  Quark-Gluon Plasma 2, Editor R.C. Hwa,  World Scientific, 1996.

\end{thebibliography}
\end{document}